# ECG-Based Blood Pressure Estimation Using Mechano-Electric Coupling Concept

SEYEDEH SOMAYYEH MOUSAVI[1], MOSTAFA CHARMI[1], MOHAMMAD FIROUZMAND[2], MOHAMMAD HEMMATI[1], MARYAM MOGHADAM[3], and YADOLLAH GHORBANI[4]

[1]Department of Electrical Engineering, University of Zanjan, Zanjan, Iran; [2]Department of Biomedical Engineering, Iranian Research Organization for Science & Technology (IROST) Tehran, Iran; [3]Department of Biomedical Engineering, Amirkabir University of Technology (Tehran- Polytechnic), Tehran, Iran; [4]Medical Equipment in Valiasr Hospital, Zanjan University of Medical Sciences, Zanjan, Iran

**Abstract** —The Electrocardiograph (ECG) signal represents the heart's electrical activity while blood pressure (BP) results from the heart's mechanical activity. Previous studies have investigated how the heart's electrical and mechanical activities are related and have referred to their relationship as the Mechano-Electric Coupling term. A new method to estimate the BP including is proposed which uses only the ECG signal. In spite of studies performed on feature extraction based on the signals' physiological parameters (Parameter-based), in this work, the feature vectors are formed with samples of the ECG signal in a particular time frame (Whole-based) and these vectors are input into Adaptive Boosting Regression (AdbootsR) to estimate BP. The nonlinear relationship which correlates BP with the ECG signal is concluded by the results of this study. According to the results, the used algorithms, for estimating both diastolic blood pressures (DBP) and Mean Arterial Pressure (MAP), are in compliance with the standards of the Association for the Advancement of Medical Instrumentation (AAMI). Also, according to the British Hypertension Society (BHS) standard, estimating DBP and MAP with the proposed method attain an A grade while it achieves B for Systolic Blood Pressure (SBP). The results indicate that using the introduced method, BP can be estimated continuously, noninvasively, without cuff, calibration-free and by using only the ECG signal.

## INTRODUCTION

As reported by the World Health Organization (WHO), 1.13 billion people worldwide suffer from Hypertension [5]. However, most of the people with this problem do not have any information about their disease. According to the fact that it does not have specific signs, it is called the silent killer [7]. In 2015, 7.5 million people died from hypertension, one of the main origins of cardiovascular diseases.

### Blood Pressure

The arteries carry blood from the heart to body organs. The blood flow in the vessels requires a proper pressure. The heart produces this pressure. Thus, the frequency of pulse pressure generation is equal to the frequency of the beating of heart. The pressure that enters the wall of the artery with any contraction of the heart muscle is called blood pressure (BP) [8]. Vascular characteristics such as elasticity and wall thickness of the vein play a significant role in forming the pressure wave propagating through the cardiovascular system of the body. BP value is amid a minimum and a maximum value called Diastolic Blood Pressure (DBP) and Systolic Blood Pressure (SBP), respectively [9]. Mean Arterial Pressure (MAP) is also defined as:

$$MAP = (2DBP + SBP)/3 \qquad (1)$$

According to the data of National Heart, Lung, and Blood Institute (NHLBI), the SBP and DBP of an adult, with a normal state, should be less than 120 mmHg and more than 80 mmHg, respectively [10]. It should be noted that these ranges vary with age and gender; for instance, SBP and DBP are in higher ranges in adults with hypertension[5].

The results of BP measurement are used to diagnose, treat and control BP related diseases such as hypertension because BP is the product of the cardiac activity and its value contains precious medical data about the cardiac activity. In addition, physicians can check the response of the cardiovascular system in the absence or presence of drugs [11].

### Cuff-less Blood Pressure Estimation

Within two recent decades, there have been conducted different studies on the basis of continuous, non-invasive and BP measurements without cuff. Utilizing pulse transit time (PTT) is one of the most well-known methods to estimate BP. PTT is referred to the propagation time of the heart beat pressure pulse, traveling from the heart to the body organs [11, 12]. Many pieces of research in the field have shown that PTT is strongly correlated with SBP and DBP which can be determined using different methods. Simultaneous Photoplethysmography (PPG) and Electrocardiograph (ECG) signals are used for measuring PTT [6]. Recently, there have been explored new methods to estimate BP based on using just PPG signals. In 2003, Teng & Zhang with the study of 15 people were able to estimate BP using only PPG signals for the first time [13]. These studies are followed by PPG signals in the time domain and transferring them to the frequency domain [14, 15]. In 2016, Xing et al. proposed a novel way of estimating BP using just the PPG signal and showed that the shape of the PPG signal and BP values have a high correlation with each other [15]. And recently, a new method based on machine learning algorithms to estimate BP using only the PPG values is proposed which estimates the BP regardless of the PPG signal's appropriate or inappropriate shape [4] and uses data obtained from sampling PPG signals.

## *Blood Pressure and Electrocardiogram Signal*

ECG signal depicts the heart's electrical activity; it shows graphically the variation in electrical potential between distinct measuring points of the body during a cardiac period. ECG signals are often recorded with means of noninvasive electrodes from the surface of skin. In hospital and medical centers, the ECG signal is used to analyze and evaluate the functioning of the heart to diagnose heart diseases or its failure. Furthermore, the applications of the ECG signal have been reported in areas such as identification [16] and respiratory rate estimation [17].

Pumping blood to the cells of the body, in order to transfer the needed material to them, is the most important task of the heart. Heart's function can be viewed from both electrical and mechanical viewpoints. The ECG waveform gives both illustration and data about the electrical activity of heart while BP gives details about the mechanical activity of heart. For explaining the activity of the heart and BP production as the result of its operation, two different views of electrical and mechanical activity should be examined which are dependent and have an interactive relationship. Mechano-Electric Coupling (MEC) term is used by cardiology researchers to explain how the electrical activity of heart is related mutually to its mechanical activity; both Excitation-Contraction Coupling (ECC) and MechanoElectric Feedback (MEF) can be justified using MEC [18]. In the last two decades, researchers have studied the interaction of the heart's mechanical and electrical parts [19].

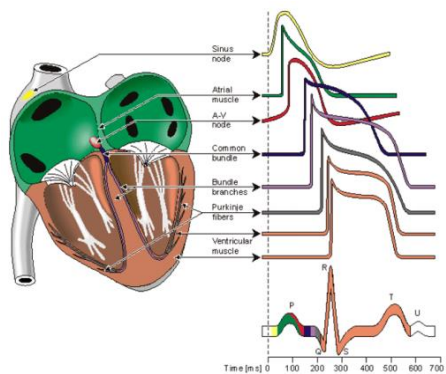

**FIGURE 1. The ECG signal [1]**

As the heart organ is formed of heart cells, in a microscopic view, the ECG signal is affected by each cell's electrical activity; this phenomenon is illustrated in Figure 1. The electrical stimulation of each heart cell causes the contraction, expansion, or mechanical work of that heart muscle cell. Therefore, the generation of pressure by the heart results in the suppression of the mechanical activity of all the cells of the heart. Of course, BP in the veins is affected by this pressure and the characteristics associated with the veins in the body.

## *Blood Pressure Estimation Using Only the ECG Signal*

In 1923, Wiggers reported the basic concepts of communication and influence of the electrical part on the mechanical part of the heart by testing them in animals [20]. In 1948, Coblentz et al. reviewed these concepts in humans. They recorded the ECG signal from the skin surface and measured BP from four different points in the body [21]. The results of their review confirmed the results of the studies of Wiggers. In general, mechanical activation and contraction of the heart requires an electrical stimulation. In other words, the electrical section of the heart controls the mechanical activity of the heart, which researchers refer to it as the Excitation-Contraction Coupling (ECC) term [22]. The influence of the electrical part of the heart on its mechanical counterpart is more explained based on chemical justifications in [23]. Figure 2 illustrates the ECC concept.

In recent years, scientists have also investigated the influence of mechanical components on the electrical part. Mechano Electrical Feedback (MEF) term refers to this connection. MEF explains electrophysiological changes in the cell scale-like decreasing the action potential duration, decreasing the diastolic action in rest and decreasing the maximum systolic action potential amplitudes [24]. The chemical concepts explaining MEF are provided in [18].

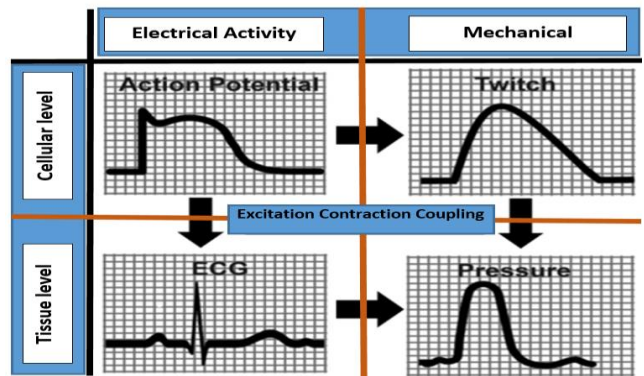

**FIGURE 2. The block diagram of the ECC concept.**

Figure 3 depicts the MEF concept.

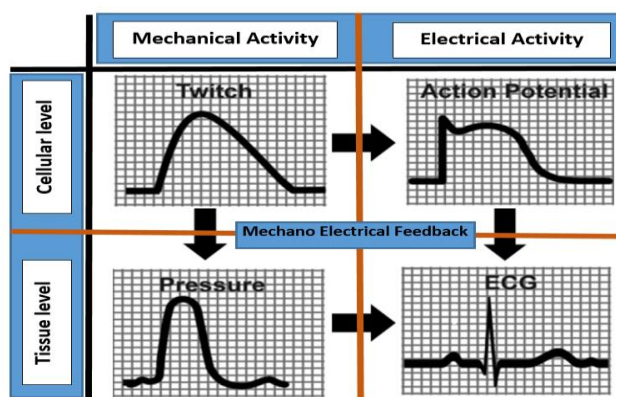

**FIGURE 3. The block diagram of the MEF concept.**

The cycle of the interaction of the mechanical and electrical parts is called the Mechano-Electric Coupling (MEC) [19]. Figure 4 illustrates the relationship between ECC and MEF.

In the field of engineering sciences, very few works have investigated BP estimation using only the ECG signal without the justification of physiological reasons relating to BP and the ECG.

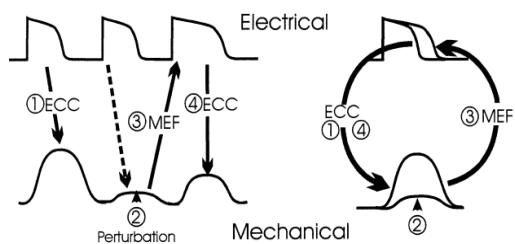

**FIGURE 4. The relationship between ECC and MEF [2]**.

In 2013, a nonlinear relationship was reported between BP and ECG signals which can be mathematically described by Taylor's series [25]. In another work which has been conducted among women climbers, it has been shown that there is a correlation between high BP and ischemic ECG signal's pattern [26]. In 2014, in a pioneering work, DBP and SBP were estimated by employing a very simple and weak classifier which labeled each of DBP and SBP only by two formerly learned values [27]. The major deficiency of this work is that its algorithm cannot be regarded as a BP estimator and its statistical population consists of only 20 people. In 2017, Yang et al. with a study of 177 people and a simple linear regression algorithm, demonstrated that the ECG voltage and SBP values are correlated in normotensive subjects which have little or no prevalence of left ventricular hypertrophy [28]. They explained that maximum pressure in the cardiac cycle is defined by stress on the myocardial fibers which can influence depolarization of membrane and cause variation of the electric field and then changes in the ECG signal acquired from a person's body surface. Moreover, in 2018, Monika et al. proposed a new algorithm which filters and then segments the untreated ECG values; afterwards, for feature extraction, it executes a complexity analysis [3]. They attained 7.72 mmHg for SBP, 8.13 mmHg for MAP, and 9.45 mmHg for DBP in terms of mean absolute error (MAE) with calibration and using data of 51 different subjects. Overall, in all of these studies, the physiological reasons relating the BP to ECG signal are not mentioned.

As explained in the above paragraph, the interaction of BP and the ECG signal can be argued with MEC concept, although it is very complicated. In studies of machine learning algorithms' applications, it is aimed to use known input-output data for future output estimation. Machine learning algorithms receive these parameters as their inputs. The parameters should be extracted, whereas the estimated values have the lowest difference with the real values. The method that uses the machine learning algorithms such that input data values are parameters based on the feature extraction of the input signals is called the Parameter-based approach [6]. One of the most important parameters is the heart rate (HR) which has individual effects on BP. Barbe et al. and Mahmood et al. used this important feature alone to estimate BP [29], [30].

## *Research Contribution*

According to studies conducted by authors of this article, the extraction of additional features from only the ECG signal has not yet been reported for better performance. In general, the Parameter-based methods which use the ECG signals have two main limitations in the estimation of BP as the following:
- ✓ Features extracted from each person's ECG signal depends on the physiological parameters of his/her body. In these methods, a general model for BP estimation is produced by machine learning algorithms. Hence, the uniqueness of the generalized model for BP estimation per person requires one or even several times of calibration. Although according to new health-care standards, it is possible to have calibration, but its removal will be very effective and useful . Still, providing a simple calibration algorithm is one of the major challenges of cuff-less BP devices [31], [6].
- ✓ To estimate BP, the features associated with it should be extracted out of the signals and then formatted to form a feature vector. There should be the highest correlation between the actual BP and BP estimator's output. No sufficient studies have been reported for estimating BP using only the ECG signal. Therefore, there is not enough information about these features.

To solve the two problems of the Parameter-based methods, in this work, a different and simple approach to form a feature vector is introduced. In this method, instead of feature extraction of the signal, the entire signal values are used at specific time intervals. This methodology which employs machine learning algorithm whose input data is a part of the raw input signal is called the Whole-based approach. For the first time, this method was reported by Kachuee et al. over the PPG signal and the PTT [6] and in a recent work this method is applied to BP estimation using only PPG signals [4]. The algorithm which is proposed in this paper is based on this method only differing in its input signal which uses the ECG signal instead of the PPG signal. Due to the fact that the ECG signal is dependent upon each individual's body, the model proposed for BP estimation in this algorithm is calibration-free.

In this research, a new method for extracting the feature vector and estimating BP using only the ECG signal continuously is proposed. This work is categorized as a noninvasive method for BP estimation. There would be no cuff used for BP estimation and calibration is not necessary for each person. In this work, using a machine learning algorithm and a signal preprocessing method, DBP, SBP and MAP values are estimated with very low mean error and appropriate standard deviation. In short, a novel method for extracting feature vectors after initial preprocessing of ECG signals is implemented. The extracted features with the so-called method are input to nonlinear adaptive boosting regression (AdaboostR) to estimate BP.

Overall, the present article is among the first studies based on physiological arguments, which BP can be estimated continuously, noninvasively, without cuff, calibration-free and using only the ECG signal. In addition, results of this work confirm the MEC.

## MATERIALS AND METHODS

The proposed algorithm of this paper is depicted in Figure 5. This block diagram consists of 3 main steps:
- ✓ Preprocessing of input data for noise reduction
- ✓ Extracting the new type (Whole-based) of feature vector
- ✓ Applying nonlinear regression algorithms to estimate BP

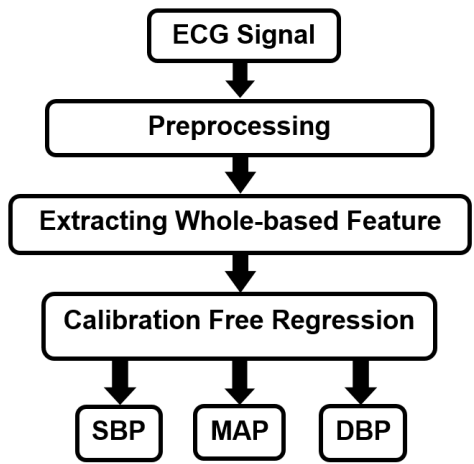

**FIGURE 5.** Proposed algorithm for BP estimation

### Database

In this work, the data source for ECG and aorta blood pressure (ABP) is selected from Multi-parameter Intelligent Monitoring in Intensive Care (MIMIC) II data (version3, 2015) [32]. Data within MIMIC II is recorded from patients in the hospital's intensive care unit. In the mentioned dataset, the ABP and the ECG signal have been recorded from the aorta in invasively and lead II, respectively [33]. There is no further useful information on this database, such as age, height, weight, and type of illness. Each dataset contains the ECG signal and BP pulse of an individual patient. The sampling of ECG and PPG signals is done simultaneously and the sampling frequency (fs) is 125 Hz. In this work, the samples are examined regarding the regularity of the HR and none of them have atrial fibrillation (AF). In AF, RR intervals are irregular, and some heartbeats may fail to make peripheral pulse. In pulse deficit, the ECG signal between RR intervals exists without corresponding BP pulse.

In each sample, the time intervals between two successive R peaks of ECG signal are approximately identical, and very small differences can be neglected. From each person's data, three distinct 15 seconds segments are separated. In this work, the minimum and maximum amounts are considered as DBP and SBP, respectively, in each ABP signal's 15 seconds interval. Then, MAP is calculated according to (1). BP estimation algorithms are applied to 1329 records which belong to 443 people. This data is for 443 people. The key statistics extracted from the database HR, DBP, MAP and SBP values are shown in Table I. Therefore, the HR outside of this range is not within the scope of this article.

**TABLE I**
**STATISTICS ON BP AND HR VALUES**

|     | Min | Max | Mean   | SD    |
| --- | --- | --- | ------ | ----- |
| **HR**  | 50  | 143 | 89.52  | 16.11 |
| **DBP** | 50  | 116 | 62.66  | 11.51 |
| **MAP** | 59  | 133 | 86.15  | 13.57 |
| **SBP** | 74  | 200 | 133.32 | 24.04 |

### Preprocessing

The noise from 50/60Hz line, baseline wander and burst noises are instances of destructive phenomena which can adversely influence the ECG signal [34]. In this paper, in a 4-step algorithm, the impurity that the noise may add to the ECG signal is eliminated. Fast Fourier Transform (FFT), Discrete Wavelet Transform (DWT), Empirical Mode Decomposition (EMD) are examples of methods for filtering data [15]. In this paper, FFT is used because of its simplicity in implementation. For filtering, the signal is transferred into frequency domain and the instances that are out of the 0.8 Hz to 40 Hz range are all zero-padded. After reconstructing signals into the time domain, the signals' amplitude is normalized.

### Extracting the Whole-based feature

As the feature vector, in this work, the entire ECG signal at a specific interval in one HR is input to the machine learning algorithm. The estimation algorithms for each of DBP, MAP, and SBP values are trained independently and separately. The feature vector includes the ECG signal values between the two consecutive R peaks. Figure 6 shows an example of feature vector extraction which is described

in the following:
- First, the feature vector is considered to be a vector with a length of 4 × fs (500 samples for this database). As stated in Table I, values of HR fall between 50 and 143. Therefore, in an ECG signal under consideration, the distance between two successive R peaks is at least 0.4 and at max 1.2 seconds.
- R peaks of the ECG signal are identified after preprocessing the signal using the Automatic Multi-scale-based Peak Detection (AMPD) algorithm [35].
- The first R peak of the ECG signal is selected as the reference peak. After choosing two consecutive R peaks, the sampled values between them are selected as the feature vector matrix. The endpoints' value may be zero in some records that their HR is more than 15 beats per minute.
- Resampling by interpolation makes very small changes in values but removes zero values

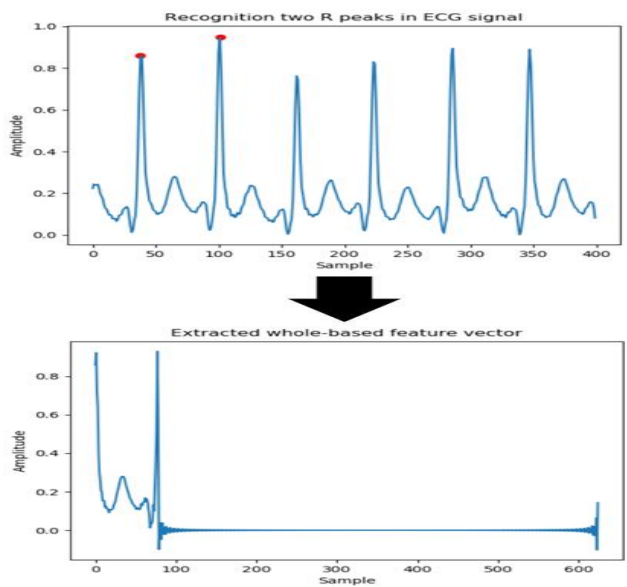

**FIGURE 6.** Extraction of the feature vector.

### *Dimension Reduction*

In the Whole-based feature vector, many features (signal values) are highly correlated with each other which this increases the need for high volumes of training data while the extracted feature vectors are too lengthy. In order to reduce the computational burden in both data training and evaluation steps, feature vectors' length can be reduced. Consequently, the feature vectors' dimension is reduced. For this purpose, the Principal Component Analysis (PCA) algorithm is utilized [36]. Implementing the mentioned process, 98% of the energy of Eigen vectors is maintained while the feature vectors are shrunk from 625 values to 17.

### *Adaptive Boosting Regression*

Based on the result of other research, AdaboostR algorithm is chosen as one of the best estimator algorithms [4]. In the AdaboostR algorithm the entire estimator is built of an array of simple estimators. Although each simple estimator produces a poor result itself in approaching the best estimation, composing weak estimators and assigning each of them a scaling weight can enhance the accuracy and performance of the entire estimator. Decision Tree Regression (DTR) algorithm is used as the simple estimator in the majority of implementations for AdaboostR. Compared to more complex algorithms, over-fitting occurs rarely in AdaboostR. As a disadvantage, this algorithm needs a huge amount of memory because of the existence of many weak DTR estimators in its structure. In addition, due to the sequential implementation of the training step, for accomplishment of the current step, the results of the previous step are required which this reduces the speed of the algorithm. Fundamentally, AdaBoostR is regarded as a meta-estimator which starts by fitting a regressor into the original dataset and then fits extra copies of the regressor on the same dataset; error of the current prediction is used as the criterion for adjusting the weights of instances. Overall, it can be stated that subsequent regressors concentrate on difficult cases more [37]. The used algorithm is implemented through the Scikit library and executed in the Python programming environment.

## RESULTS

As mentioned before, the results of the extraction of feature vectors of the previous stage are considered as nonlinear regression inputs. At this nonlinear regression model, a k-fold cross validation (CV) technique is used. To examine the capability of a feature vector in BP estimation, the 10-fold CV is performed which separates the data associated in learning and testing. In Figure 7, Bland Altman plots for DBP, SBP, and MAP values in one of the folds are provided; it can be observed that the vast majority of the estimated values are settled in the vicinity of lines and their deviation from the target line is less than 5 mmHg.

For evaluating the accuracy of an algorithm for BP estimation, there are many standards. The AAMI standard assesses an algorithm based on the ME and the SD of the errors; an algorithm which its ME and SD are less than 5 and 8 mmHg, respectively, satisfy the boundaries of this standard [38]. The BHS standard also evaluates the algorithms based on the cumulative percentage of absolute errors in three different categories 5, 10 and 15 mmHg [39]. Based on both standards, an algorithm should be run at least on 225 samples. The AdaboostR algorithm of this article has been run on 1329 samples. In Table II, the comparison of the results of AdaBoostR algorithm is shown in BHS standard. Based on BHS, AdaBoostR algorithm achieves an A grade for DBP and MAP and B for the estimating SBP. Checking the algorithm results with AAMI standard is available in Table III. The accuracy of AdaboostR is completely accepted for DBP and MAP estimations by AAMI standard, and the results are in the close vicinity of the standard boundary for SBP estimation.

## DISCUSSION

First, this article explains that there is a relationship which correlates BP into ECG signal based on physiological arguments expressed by MEC, ECC, and MEF concepts.

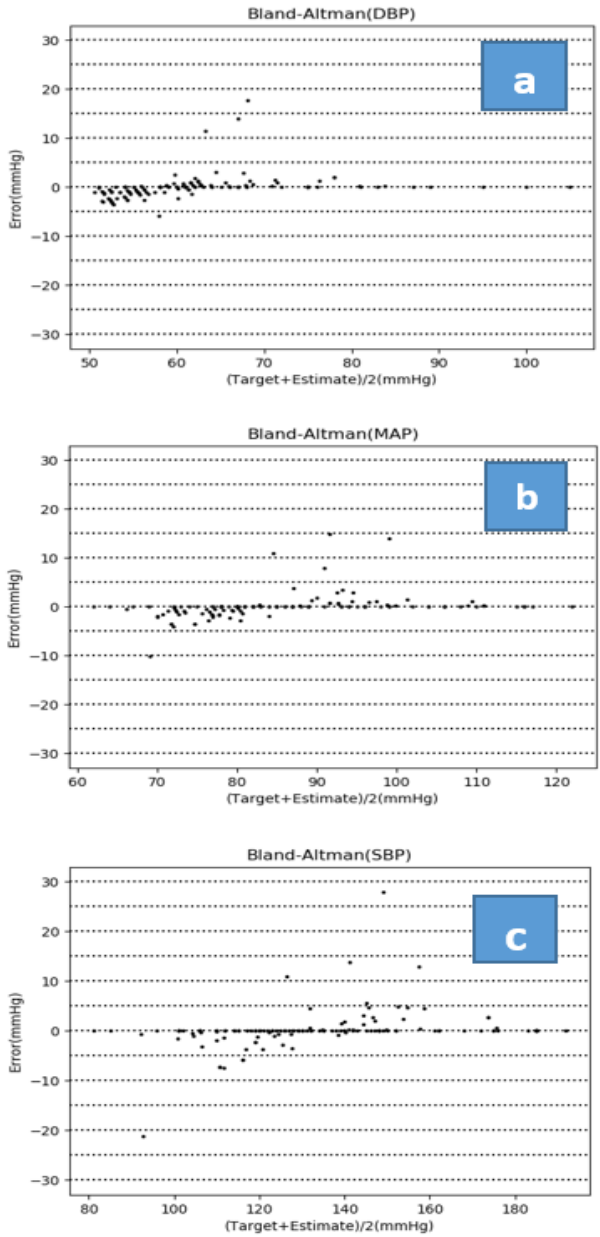

**FIGURE 7.** Bland-Altman for: a) DBP, b) MAP, and c) SBP

**TABLE III**
**The Comparison of the Results of This Paper with The AAMI Standard**

|  |  | Mean (mmHg) | SD (mmHg) | Subject |
|---|---|---|---|---|
| This Work | DBP | -0.100 | 4.201 | 443 |
|  | MAP | -0.154 | 4.629 | 443 |
|  | SBP | -0.291 | 8.831 | 443 |
| AAMI | BP | ≤5 | ≤8 | ≥85 |

Then, a novel continuous and noninvasive BP estimation method is proposed which not only removes cuff but it also does not need any calibration; besides, it uses only the ECG signal.

Although only a few studies are released on measuring blood pressure using only the ECG signals, the achievements of the current study are compared to the eminent works within this topic in Table IV which they have the best results among many other methods.

The proposed methodology of the current paper excels the one presented in [3] in terms of accuracy for estimation of SBP, DBP, and MAP. These results emphasize appropriate performance of the Whole-based method compared to Parameter-based method; because the feature vector in Whole-based method includes features that have high correlation with BP values. In addition, the proposed method has better performance rather than [4] which estimates BP by using the PPG signal by Whole-based method and also [6] which estimates BP by using only PPG signals and the PTT concept by Whole-based method.

Medical reports reveal that hypertension is regarded as a significant motive of diseases related to heart or blood vessels (cardiovascular). Control and diagnosis of this disease is done via continuous monitoring of BP. In recent years, measurement of BP without cuff has been investigated, which record simultaneous PPG and ECG signals, and accordingly, PTT can be measured. Subsequently, studies on the association of only the PPG signal and BP are studied and modeled, so that the results of the calculation of BP are reported using only the PPG signal. Furthermore, the cardiology researchers have conducted extensive studies to figure out and explain how mechanical and electrical activities of the heart are related. They have shown that these two parts affect each other and their results are reported with the concepts of MEC, MEF, and ECC. Therefore, it can be argued that the calculation of BP is justifiable by only ECG signals.

In this paper, the physiological concepts relating ECG and BP signals are reviewed; based on the mentioned concepts, BP is estimated using just the ECG signal, by an effective Whole-based algorithm; this method is noninvasive and without cuff. The feature extraction algorithm uses sampled ECG values which are raw. Furthermore, in this algorithm executing calibration steps for each individual is not necessary.

To the best knowledge of authors, estimating BP using only ECG signals have not been comprehensively investigated so far. The first study [27] within this topic has very simple and weak algorithm that is not an estimator but a classifier which assigns two different values to each of SBP and DBP. In this paper, for reporting results of error, the difference between real values and one of the associated labels are calculated. Therefore, results have not been reported based on standards for BP estimation equipment. Hence, comparing the results of them with the performance of the algorithm proposed in this work is impossible. The second study [3] in this topic is performed on few subjects (51) and needs a calibration step. Furthermore, in this study, the physiological reasons relating the ECG signal to the BP are not covered.

The proposed algorithm has a very powerful estimator. Also, the statistical population of this study is sufficiently large and diverse including a database of 443 individuals which have regular HR and do not have atrial fibrillation.

This study confirms the existence of the so-called nonlinear relationship. Moreover, the results demonstrate the feasibility of estimating BP using just the ECG signal. According to our results, the introduced solutions attain an A grade for estimating DBP and MAP and B for the SBP based on BHS standard. Also, according to AAMI standard, the results of the proposed algorithms are completely in the acceptable range of standard for both DBP and MAP estimations and are very close to the standard boundary for SBP estimation.

According to the diversity and a large amount of data used in training and testing and also not using individual calibration step, the results affirm that the proposed algorithm is robust. It is anticipated that if healthy people's data were added to the MIMIC database, there would undoubtedly be significant improvements in the accuracy of the proposed algorithm. Overall, this method can be used in different applications like exercise test. Furthermore, it can be used in building ECG Holters which in such equipment, ECG signal is recorded for 24 to 48 hours during daytime activities and night-time sleep. Consequently, the patient's BP will be accessible to physicians with an acceptable precision, without pain and without any invasive action

### TABLE II
### Comparison Results Of The Current Study with The BHS Standard

| | | Cumulative Error Percentage | | |
|---|---|---|---|---|
| | | ≤5mmHg | ≤10mmHg | ≤15 mmHg |
| **Results** | DBP | $\frac{1208}{1329} \times 100 = 90\%$ | $\frac{1296}{1329} \times 100 = 97\%$ | $\frac{1329}{1329} \times 100 = 100\%$ |
| | MAP | $\frac{1174}{1329} \times 100 = 88\%$ | $\frac{1264}{1329} \times 100 = 95\%$ | $\frac{1329}{1329} \times 100 = 100\%$ |
| | SBP | $\frac{1120}{1329} \times 100 = 84\%$ | $\frac{1198}{1329} \times 100 = 90\%$ | $\frac{1242}{1329} \times 100 = 93\%$ |
| **BHS** | Grade A | 60% | 85% | 95% |
| | Grade B | 50% | 75% | 90% |
| | Grade C | 40% | 65% | 85% |

### TABLE IV
### Comparing Achievements Of The Current Study with The Recent Research

| Study | Source | Subject | DBP | | MAP | | SBP | |
|---|---|---|---|---|---|---|---|---|
| | | | MAE | SD | MAE | SD | MAE | SD |
| **This work** | ECG | 443 | 1.94 | 3.72 | 2.04 | 4.15 | 3.66 | 8.06 |
| [3] | ECG | 51 | 7.72 | 10.22 | 9.45 | 10.03 | 8.13 | 8.84 |
| [4] | PPG | 443 | 2.43 | 3.37 | 2.61 | 4.16 | 3.97 | 7.99 |
| [6] | PPG and ECG (PTT) | 942 | 6.14 | 5.35 | 5.38 | 5.92 | 10.06 | 11.17 |

## CONFLICT OF INTEREST
There are no conflicts of interest.

## REFRENCES